\documentclass[journal,twoside,web]{ieeecolor}
\usepackage{generic}
\usepackage{cite}
\usepackage{amsmath,amssymb,amsfonts}
\usepackage{algorithmic}
\usepackage{graphicx}
\usepackage{textcomp}
\usepackage[linesnumbered,ruled,vlined,boxed,commentsnumbered]{algorithm2e}
\usepackage{hyperref}
\usepackage{subcaption}

\newtheorem{lemma}{Lemma}
\newtheorem{assumption}{Assumption}

\newtheorem{cor}{Corollary}
\newtheorem{remark}{Remark}

\newcommand{\norm}[1]{\left\lVert#1\right\rVert}

\def\BibTeX{{\rm B\kern-.05em{\sc i\kern-.025em b}\kern-.08em
    T\kern-.1667em\lower.7ex\hbox{E}\kern-.125emX}}
\begin{document}
\title{Koopman based Data-driven Simulation and Control}
\author{Yingzhao Lian$^\dagger$, Renzi Wang$^\dagger$, and Colin N.Jones$^\star$
\thanks{This work has received support from the Swiss National Science Foundation under the RISK project (Risk Aware Data-Driven Demand Response), grant number 200021 175627  }
\thanks{$^\dagger$: The first two authors contributed equally to this work.}
\thanks{$^\star$: Corresponding author }
\thanks{Yingzhao Lian, Renzi Wang and Colin N. Jones are with Automatic Laboratory, Ecole Polytechnique Federale de Lausanne, Switzerland. \tt\small $\{$yingzhao.lian, renzi.wang, colin.jones$\}$@epfl.ch}}


\maketitle

\begin{abstract}
Sparked by the Willems' fundamental lemma, a class of data-driven control methods has been developed for LTI system. At the same time, the Koopman operator theory attempts cast a nonliner control problem into a standard linear one albeit infinite dimensional. Motivated by these two ideas, a data-driven control scheme for nonlinear systems is proposed in this work. The proposed scheme is compatible with most differential regressor enabling an offline learning. In particular, the model uncertainty is considered, enabling a novel data-driven simulation framework based on Wasserstein distance. Numerical experiments are performed with Bayesian neural networks to show the effectiveness of both the proposed control and simulation scheme.

\end{abstract}

\begin{IEEEkeywords}
Predictive control, data-driven control, Koopman operator
\end{IEEEkeywords}

\section{Introduction}
\label{sec:introduction}
\IEEEPARstart{R}{ecent} trend of the digitalization motivates the research interest of data-driven control~\cite{hou2013model} because of the wide access of data collected by the sensors. Instead of resorting to a first-principle model, the collected data are used either to identify a model~\cite{ljung1999system} or to construct a controller directly. The former approach is compatible with most control theory and therefore results in successful applications~\cite{lanzetti2019recurrent,kocijan2016modelling}. Featuring a controller desgin without any intermediate stage, the letter data-driven scheme attract more research interests and finds successful application in linear systems~\cite{campi2002virtual} and iterative control~\cite{hjalmarsson2002iterative,bristow2006survey}. It worth mentioning that in the community of reinforcement learning~\cite{sutton2018reinforcement}, the learning schemes can also be categorized as model-based methods and model-free methods.

This work is developed on the basis of the Willems' fundamental lemma and the Koopman operator theory. In particular, the Willems' fundamental lemma characterizes the responses of deterministic linear time invariant(LTI) systems with measured trajectories under reasonable assumptions of controllability and persistent excitation. Based on the data-driven prediction enabled this lemma, predictive control scheme has been developed~\cite{coulson2019data,markovsky2007linear}. Beyond the predictive control, the Willems' fundamental lemma has also been adopted in feedback controller design~\cite{berberich2020robust,de2019formulas}. Within the LTI framework, the Willems' fundamental lemma is further extend to incorporate measurement noise~\cite{9103015,yin2020maximum} and process noise~\cite{berberich2020robust}. 

At the same time, a significant collection of works has tried to extend the applications of the Willems' fundamental lemma to nonlinear systems. In~\cite{berberich2020trajectory}, an extension to Hammerstein systems and Wiener systems is proposed based on an a priori knowledge of basis function. \cite{bisoffi2020data,rueda2020data,guo2020data} attempt to apply the Willems' fundamental lemma to the class of polynomial system. As pointed out in~\cite{rueda2020data}, the necessary and sufficient condition in the fundamental lemma is broken. An promising viewpoint regarding the quotient space is proposed in~\cite{lian2021nonlinear}, which clusters trajectories into equivalent class and extend the Willems' fundamental lemma into the reproducing kernel Hilbert space. The functional space viewpoint in~\cite{lian2021nonlinear} motivates us to apply the Willems' fundamental lemma in the function space. In particular, the Koopman operator theory is used.

The Koopman operator theory is first introduced in the study of forward-complete autonomous sytems~\cite{Koopman255,Koopman315}, which is a linear composite operator even when the system is nonlinear. \cite{korda2018linear,surana2016linear} later introduced the applications of the Koopman operator in controller and observer design, followed by a wide range of research ranging from the model reduction~\cite{peitz2019koopman} to the global optimal control~\cite{villanueva2020towards}. Even though most algorithms based on the Koopman operator have numerical implementations that are similar to those studied in~\cite{berberich2020trajectory,bisoffi2020data,guo2020data,rueda2020data}, the Koopman operator establishes a totally different theoretical framework. In particular, the Koopman operator corresponds to a Heisenberg picture which models the evolution of the observable, while other aforementioned methods model the evolution of the state, corresponding to a Schrödinger picture~\cite{landau2013quantum}. The Koopman operator theory can alleviate the theoretical issue in studying the lifting function without resorting to the quotient space, and enables convergence analysis as that has done in~\cite{korda2018convergence}. 

The key component of a Koopman operator based method is the learning of the eigenfunctions or the lifting functions lying within the subspace spanned by the eigenfunctions. A standard framework of extended dynamic mode decomposition (EDMD) spans the lifting functions with a dictionary of basis function~\cite{williams2015data}, which suffers from the curse of dimensionality. To overcome this challenge, \cite{kawahara2016dynamic,klus2020eigendecompositions} applies kernel method to learn the Koopman operator in a non-parametric way, which is still not scalable to large dataset. In \cite{takeishi2017learning,lusch2018deep}, the lifting functions are approximated by neural networks. However, these aforementioned methods mainly consider one-step forward prediction either due to the formulation of the learning problem or due to numerical stability, which results in a relatively inaccurate long-term prediction. A link between the Koopman operator and the subspace identification is observed in~\cite{lian2019learning}, which enables a learning scheme of long-term prediction. However, this method still suffers from the lack of scalability and the numerical limitation of the subspace identification~\cite{qin2006overview}.

In this work, we propose to incorporate the learning of a Koopman operator into the framework of the Willems' fundamental. The applications of Koopman operator in the Willems' fundamental lemma are mentioned in~\cite{coulson2019data,berberich2020trajectory} but have not been detailed. In this work, we show that by maximizing the linearity of an finite order approximation, a Koopman operator can be learned based on the sensitivity analysis of a parametric programming problem. The proposed learning scheme is capable of uncertainty quantification, where a new objective function is derived to account for the uncertainty. Meanwhile, we propose a control scheme that solves a bi-level optimization problem by a transformation into to a single level structure. The bi-level optimization formulation has been discussed in both~\cite{coulson2019data,dorfler2021bridging}, where a bi-level problem is relaxed to a multi-objective problem. 

The remainder of this paper is organized as follows: the preliminary knowledge is introduced in Section~\ref{sec: pre}, after which the training and the prediction based on the proposed scheme is elaborated in Section~\ref{sec: prediction}. We explain the proposed control framework in Section~\ref{sec: control} along with presenting the numerical simulation results of prediction and control in Section~\ref{sec: simulation}.


\section*{Notations}
$\norm{x}_p$ indicates the $\ell_p$ norm of vector $x$ and $\norm{x}_Q := x^\top Q x$ is the weighted norm with $Q$ being positive semi-definite. $\norm{A}_F$ and $\norm{A}_*$ denote the Frobenius norm and the nuclear norm of the matrix $A$ respectively. $\mathcal{N}\sim(\mu, \Sigma)$ is a Gaussian distribution with mean value $\mu$ and covariance matrix $\Sigma$. We use $\mathbb{Z}_{\geq 0}$ to represent a non-negative integer. $\textbf{w}:=\{w_k\}_{k=a}^{b}$ is a sequence of signal $\{w_a, \dots, w_b\}$ indexed by $k$. Specifically, the boldface is used to denote a sequence while the lightface denotes a measurement, \textit{e.g.} $\textbf{w}$ and $w_k$. Meanwhile, the subscript $d$ is reserved to denote the data collected offline. The superscript $^*$ is used to denote optimal solution of an optimization problem. $\otimes$ denotes the Kronecker product.

\section{Preliminary}\label{sec: pre}
In this section, the Willems' fundamental lemma and the Koopman operator theory will be introduced. Then, the sensitivity analysis of a parametric optimization, which is the enabler of the learning, is discussed.
\subsection{Willems' Fundamental Lemma}
Given a sequence of measurements $\{w_k\}_{k=0}^{T-1}$, its Hankel matrix of depth $L$ is defined as 
\begin{equation}\label{eq:hankel}
    H_L(\textbf{w}) := \begin{bmatrix}
    w_0 & w_1 & \dots & w_{T-L}\\
    w_1 & w_2 & \dots & w_{T-L+1}\\
    \vdots & \vdots & \ddots & \vdots \\
    w_{L-1} & w_{L} & \dots & w_{T-1}
    \end{bmatrix}\;.
\end{equation}
Regarding a Hankel matrix $H_L(\textbf{w})$, the signal sequence $\textbf{w}$ is persitently exciting of order $L$ if $H_L(\textbf{w})$ is full row rank. The Willems' fundamental lemma utilizes the Hankel matrices to characterize the response of the following deterministic linear time invariant(LTI) system, dubbed $\mathfrak{B}(A,B,C,D)$,
\begin{equation}\label{eq: linear_dynamics}
\begin{aligned}
        x_{k+1}&=Ax_k+Bu_k\\
        y_k &= Cx_k+Du_k
\end{aligned}\;, 
\end{equation}
where $A\in\mathbb{R}^{n_x\times n_x}, B\in\mathbb{R}^{n_x\times n_u}, C\in\mathbb{R}^{n_y\times n_x}, D\in\mathbb{R}^{n_y\times n_u}$ parametrize the system dynamics and the order of this system is denoted by $O(\mathfrak{B}(A,B,C,D)):=n_x$. The \textbf{Willems' fundamental lemma} is concluded as

\begin{lemma}\label{lemma: fundamental lemma}
(\cite[\textit{Theorem 1}]{willems2005note}, \cite[\textit{Lemma 2}]{de2019formulas})
Consider a controllable and observable system (\ref{eq: linear_dynamics}), if the input sequence $\textbf{u}_d=\{u_{d,k}\}_{k=0}^{T_d-1}$ is persistently exciting of order $O(\mathfrak{B}(A,B,C,D)) + L$, then
\begin{enumerate}
    \item Any $L$-step input/output trajectory of system (\ref{eq: linear_dynamics}) can be expressed as
    \begin{equation*}
    \begin{bmatrix}
        H_L(\textbf{u}_d) \\ H_L(\textbf{y}_d)
    \end{bmatrix} g = 
    \begin{bmatrix}
        u \\ y
    \end{bmatrix}
    \end{equation*}
    
    \item Any linear combination of the columns of the Hankel matrices, that is
    \begin{equation*}
        \begin{bmatrix}
        H_L(\textbf{u}_d) \\ H_L(\textbf{y}_d)
    \end{bmatrix} g
    \end{equation*}
    is a $L$-step input/output trajectory of (\ref{eq: linear_dynamics})
\end{enumerate}
\end{lemma}

This lemma enables data-driven simulation and control~\cite{markovsky2008data,coulson2019data}. To make an $N$-step prediction, the Hankel matrices composed of offline data is partitioned as
\begin{equation*}
    \begin{aligned}
    \begin{bmatrix}
        U_p \\ U_f
    \end{bmatrix} := H_{T_{ini} + N}(\textbf{u}_d)\;,\; 
    \begin{bmatrix}
        Y_p \\ Y_f
    \end{bmatrix} := H_{T_{ini} + N}(\textbf{y}_d)\;,\; 
    \end{aligned}
\end{equation*}
where the first $T_{ini}$ row blocks are used to construct $U_p\;,Y_p$ while the remaining row blocks is assigned to $U_f\;,\;Y_f$. In the remainder of this paper, $n_c$ is reserved to denote the number of columns in the Hankel matrix. In particular, $T_{ini}$ is chosen to ensure the uniqueness of prediction and the rank of the observability matrix
\begin{align*}
    \mathcal{O}_{T_{ini}}(A,C) := \begin{bmatrix}
        C^\top& (CA)^\top, &\dots,& (CA^{T_{ini}-1})^\top  
    \end{bmatrix}^\top\;
\end{align*}
is of rank $O(\mathfrak{B}(A,B,C,D)=n_x$~\cite{markovsky2008data}. Without measurement noise, the  $N$-step output prediction $\textbf{y}$ is defined by
\begin{align}\label{eq:lin_pred}
\begin{split}
        \textbf{y} &= Y_fg\\
     \text{s.t.}\;\;  \begin{bmatrix}
        U_p\\ Y_p \\ U_f
    \end{bmatrix} g&=
    \begin{bmatrix}
        \textbf{u}_{ini}\\ \textbf{y}_{ini} \\\textbf{u}
    \end{bmatrix}\;,
\end{split}
\end{align}
where $\textbf{u}_{ini}$ and $\textbf{y}_{ini}$ are $T_{ini}$-step previous measurements of the inputs and the outputs. Accordingly, $\textbf{y}$ is the $N$-step response driven by input sequence $\textbf{u}$. Built on this prediction scheme, the data-enabled predictive control(DeePC)~\cite{coulson2019data} is
\begin{equation}\label{eq: DeePC}
\begin{aligned}
    \min_{g, \sigma_y, u, y} &(\sum_{k=0}^{N-1}\norm{y_k-r_{t+k}}_Q^2+\norm{u_k}_R^2)\\
    & +\lambda_g\norm{g}_1+\lambda_y\norm{\sigma_y}_1\\
    \text{s.t.} & \begin{bmatrix}
        U_p\\ Y_p \\ U_f \\ Y_f
    \end{bmatrix} g=
    \begin{bmatrix}
        \textbf{u}_{ini}\\ \textbf{y}_{ini} \\\textbf{u} \\ \textbf{y}
    \end{bmatrix} + 
    \begin{bmatrix}
        0 \\ \sigma_y \\0 \\0
    \end{bmatrix}\\
    & u_k \in \mathcal{U}, \forall k \in {0, \dotsc, N-1}\\
    & y_k \in \mathcal{Y}, \forall k \in {0, \dotsc, N-1}\;,
\end{aligned}
\end{equation}
where $Q$ and $R$ are the weight penalizing outputs and inputs respectively and $\sigma_g$ is introduce to deal with measurement noise. $r\in \mathbb{R}^{pN}$ is the reference trajectory and $\mathcal{U,Y}$ are the feasible sets of inputs and outputs. $\norm{g}_1$ and $\norm{\sigma_y}_1$ are regularization terms. $\lambda_y, \lambda_g \in \mathbb{R}_{>0}$ are regularization parameters. These regularization terms have been interpreted under distributionally robust optimization framework~\cite{coulson2019regularized} and maximal likelihood framework~\cite{yin2020maximum}.

\textit{Remark}: When a long input-output sequence is not available, the Hankel matrix $H_L(\textbf{w})$ can be replaced by a mosaic Hankel matrix\cite{van2020willems}. Given $M$ trajectories:
\begin{equation*}
\begin{aligned}
    &\textbf{w}=[\textbf{w}_1,\dotsc,\textbf{w}_M], \\
    \text{where each trajectory } &\textbf{w}_i=(w_{i,1},\dotsc,w_{i,T_i}), w_{i,k}\in\mathbb{R}^q
\end{aligned}
\end{equation*}
the mosaic Hankel matrix is defined as :
\begin{equation*}
    H_L(\textbf{w}) = [H_L(\textbf{w}_1), \dots, H_L(\textbf{w}_M)]
\end{equation*}

\subsection{Koopman Operator}\label{sec:koopman}
Given a discrete-time autonomous system
\begin{align}\label{eq:nonlin_dyn}
    x_{k+1} = f(x_k)\;,
\end{align}
where $f$ models the nonlinear dynamics, a Koopman operator is a composite operator
\begin{align*}
    \mathcal{K}\psi := \psi\circ f\;,
\end{align*}
which $\psi: \mathbb{R}^{n_x}\rightarrow \mathbb{R}$ is called observable. Unlike standard state space model, the Koopman operator models the evolution of a function driven by system dynamics $f$ and its existence is guaranteed for forward-complete system~\cite{bittracher2015pseudogenerators}. As the Koopman operator is an operator on a function space, $\mathcal{K}$ is in general infinite-dimensional, but critically it is linear even when the dynamics F are non-linear and as such, an observable $\phi$ is an eigenfunction associated with the eigenvalue $\lambda\in\mathbb{C}$ if $\mathcal{K}\phi = \lambda\phi$. From this we can see that the eigenfunctions (or linear combinations of the eigenfunctions) evolve linearly along the trajectories of our nonlinear system~\eqref{eq:nonlin_dyn}
\begin{align}
    \phi(x_{k+1})=\phi(f(x_k))=(\mathcal{K}\phi)(x_k)=\lambda\phi(x_k)\;.
\end{align}
Given a collection of eigenfunctions $\{\phi_i\}_{i=1}^{n_\phi}$, any observable lying within the span of these eigenfunctions can be decomposed into $\psi = \sum_i c_i(\psi)\phi_i$, where $c_k(\psi)$ is called the Koopman modes of $\psi$. Then, we have
\begin{align*}
    \mathcal{K} \psi = \sum_i c_i(\psi)\lambda_i\phi_i\;,
\end{align*}
with $\lambda_i$ denoting the eigenvalue of $\phi_i$.

In the sequel, the subscript $_u$ is used to denote the components corresponding to a system with control inputs. Given a nonlinear dynamics with control input
\begin{align*}
    x_{k+1} = f_u(x_k,u_k)\;,
\end{align*}
the Koopman operator can be defined in different ways~\cite{williams2016extending,proctor2018generalizing,korda2018linear}. In this work, we consider the framework in~\cite{korda2018linear}. More specifically, denote the infinite control sequence $\boldsymbol{u}:= \{u_k\}_{k=0}^{\infty}\in\mathit{l}(\mathcal{U})$, where $\mathit{l}(\mathcal{U})$ represents the space of all control sequence. The augmented state is
\begin{align*}
    \chi = \begin{bmatrix}
        x\\\boldsymbol{u}
    \end{bmatrix}\;,
\end{align*}
upon which the system dynamic is augmented as $F: \mathbb{R}^{n_x}\times \mathit{l}(\mathcal{U})\rightarrow \mathbb{R}^{n_x}\times \mathit{l}(\mathcal{U})$
\begin{align}\label{eq:Koopman_control}
    F(\chi_k) = \begin{bmatrix}
        f_u(x_k,\boldsymbol{u}_k(0))\\\mathcal{S}\boldsymbol{u}_k
    \end{bmatrix}\;.
\end{align}
$\mathcal{S}$ is the left shift operator with $\mathcal{S\boldsymbol{u}}(i):= \boldsymbol{u}(i+1)$ and $\boldsymbol{u}(i)$ is the evalution of the $i$-th element of $\boldsymbol{u}$. In this setup, $\boldsymbol{u}$ can be considered as a sequence of mappings from index $i$ to actual output $u_i$. It is noteworthy to point out that this dynamical system~\ref{eq:Koopman_control} is infinite dimensional but autonomous. Hence, the aforementioned definition of the Koopman operator can be applied directly and the corresponding eigenfunctions are assumed to be spanned by the following dictionary of basis functions.
\begin{align*}
    \{\phi_u(x,\boldsymbol{u})\}_{i=1}^{n_{\phi_u}+n_u} := \{\phi_{u,1}(x),\dots,\phi_{u,n_{\phi_u}}(x),\boldsymbol{u}(0)^\top\}\;.
\end{align*}

If the evolution of this dictionary of basis functions is closed under the system dynamics, then we have 
\begin{align}\label{eq:lifted_space_state}
\begin{split}
    z_{k+1} &= \mathcal{A}z_k+\mathcal{B} \boldsymbol{u}_k(0)\\
    \boldsymbol{u}_{k+1}(0) &= \boldsymbol{u}_k(1)\;,
\end{split}
\end{align}
where $z_k:=[\phi_{u,1}(x_k),\dots,\phi_{u,n_{\phi_u}}(x_k)]$ and $\mathcal{A,B}$ captures the Koopman operator. Similarly, any functions within the span of these basis functions can be recovered by the Koopman mode as
\begin{align}
    \psi_u(x,\boldsymbol{u}(0)) = c_u^\top \begin{bmatrix}
        z\\\boldsymbol{u}(0)
    \end{bmatrix}\;,
\end{align}
with $c:=[c_{u,1},c_{u,2}\dots,c_{u,n_{\phi_u}+n_u}]$ the vector of Koopman modes. In particular, we are interested in the Koopman mode of the identity functions evaluated on the sytem outputs. Assumed that we have $n_y$ outputs, the evaluation of the $i$-th output is $I_{y,i}(y_k):= y_{k,i}$. With a bit of abuse of notation, the Koopman modes decomposition of outputs evaluation is 
\begin{align}\label{eq:lifted_sapce_output}
     y_k = \begin{bmatrix}
         I_{y,1}(y_k)\\I_{y,2}(y_k)\\\vdots\\I_{y,n_y}(y_k) 
     \end{bmatrix}= \begin{bmatrix}
             c_{u,1}^\top\\c_{u,2}^\top\\\vdots\\c_{u,n_y}^\top
         \end{bmatrix}\begin{bmatrix}
        z\\\boldsymbol{u}(0)
    \end{bmatrix} := C_u\begin{bmatrix}
        z\\\boldsymbol{u}(0)
    \end{bmatrix}\;,
\end{align}
where $C_u$ stacks the Koopman modes of the output evaluations.

\subsection{Differential Parametric Optimization}\label{sec:diff_opt}
Sensitivity analysis investigates the smoothness of a parametric optimization problem, where the implicit function theorem~\cite{krantz2003introduction} is applied to the KKT system. This idea has been applied to deep learning~\cite{el2019implicit} and reinforcement learning~\cite{zanon2020safe}.  Though the solution map is barely differentiable, the optimal value function is smoother than the solution map~\cite{fiacco2020mathematical}, which is the only tool used in this work. In general, the continuity of a general convex optimization problem is guaranteed by the uniform level boundness~\cite[Theorem 1.17]{rockafellar2009variational}, while a general nonlinear parametric optimization problem guarantees a lower semiconitinous value function under the assumption of local compactness~\cite{bank1982non}. 

For the sake of clarity, we elaborate this derivative by a standard quadratic programming (QP), please refer to~\cite{agrawal2019differentiating} for a general conic form. We use subscript $_q$ to avoid confusion. Consider the a parametric QP, $\mathcal{Q}(e_q):=e_q\rightarrow z_q^*$ with parameters $\{Q_q,\;,q_q\;H_q,\;h_q,\;E_q\}$ and $Q_q$ positive definite:
\begin{equation}
\begin{split} \label{eqn:quaratic_program}
\underset{z_q}{\min}\ &\frac{1}{2}z_q^TQ_qz_q + q_q^Tz
\\
\text{s.t.}\ & Hz_q \leqslant h_q, E_qz_q=e_q
\end{split}
\end{equation}
The KKT conditions for the QP are:
\begin{equation}
\begin{split} \label{eqn:KKT_conditon}
    Q_qz^* + q_q + H_q^T\lambda^* + E_q^T\nu^*  &= 0
    \\
    \text{diag}\left( \lambda^* \right) (H_qz^*-h_q) &= 0
    \\
    E_qz_q^* - e_q & = 0
\end{split}
\end{equation}
where $z_q^*, \nu^*, \lambda^*$ are the optimal primal and dual variables, $\text{diag}(x)$ builds a diagonal matrix composed of $x$. Then the differentials of KKT conditions can be computed as:
\begin{equation} \label{eqn:KKT_diff}
\begin{split}
    \left[\begin{array}{ccc}
        Q_q & H_q^T & E_q^T  \\
        D(\lambda^*)A & \text{diag}(H_qz^*-h_q) & 0 \\
        E_q & 0 & 0
    \end{array} \right]
    \left[ \begin{array}{c}
         dz_q \\
         d\lambda \\
         d\nu
    \end{array}\right] \\
    = -\left[ \begin{array}{c}
         dQ_qz_q^* + dq_q + dH_q\lambda^* + dE_q^T\nu^* \\
         \text{diag}(\lambda^*)dH_qz_q^* - \text{diag}(\lambda^*)db \\
         dE_qz^* - de_q
    \end{array}\right]
\end{split}
\end{equation}
The derivatives of $z^*$ with respect to the parameters ($Q_q,q_q,H_q,h_q,E_q$) and the function input $f$ are given by the solution to the linear system defined in Equation~\eqref{eqn:KKT_diff}. For example, the solution $dz$ of~\eqref{eqn:KKT_diff} gives the result of $\frac{\partial{z_q^*}}{\partial{Q_q}}$ if we set $dQ_q=I$ and the differentials of other parameters to 0. The gradient of optimal value $L(z^*)$ with respect to $Q$ is calculated accordingly  as $\frac{\partial{L(z_q^*)}}{\partial{z_q^*}}\frac{\partial{z_q^*}}{\partial{Q_q}}$.

\section{Koopman based Data-driven Prediction}\label{sec: prediction}
In this section, the fundamental lemma is first introduced in the Koopman operator theory, which enables a training scheme by minimizing the prediction error with respect to the training dataset. The stochastic prediction scheme is thereby introduced to show the compatibility of probabilistic models, such as Bayesian neural networks and Gaussian processes. 

\subsection{Koopman Operator with the Fundamental Lemma}
As discussed in Section~\ref{sec: pre}, the key component of a Koopman operator is the eigenfunctions or the linear subspace containing the eigenfunctions. Therefore, the learning of an Koopman operator is equivalent to find functions whose evolution of the function evaluation behaves like a linear system~\ref{eq: linear_dynamics}. Following corollary is the enabler of the proposed learning scheme
\begin{cor}\label{cor:lin}
An dynamical system of order $n_x$ can be parametrized as a linear system~\eqref{eq: linear_dynamics} if and only if the Fundamental lemma holds.
\end{cor}
\begin{proof}
Necessary condition holds by Lemma~\ref{lemma: fundamental lemma} and the sufficient condition holds by the definition of linear systems.
\end{proof}

As discussed in Section~\ref{sec:koopman}, the outputs evaluations are assumed to be spanned by the following basis functions
\begin{align*}
    \{\phi_u(x,\boldsymbol{u})\}_{i=1}^{n_{\phi_u}+n_u} := \{\phi_{u,1}(x),\dots,\phi_{u,n_{\phi_u}}(x),\boldsymbol{u}(0)^\top\}\;.
\end{align*}
Then given a sequence of state evolution $\textbf{x}_d $ with its corresponding inputs-output sequence $\textbf{u}_d,\textbf{y}_d$, whose inputs are persistently excited of order $n_{\phi_u}-n_u+L$, Corollary~\ref{cor:lin} implies that $\{\phi_u(x,\boldsymbol{u})\}_{i=1}^{n_{\phi_u}+n_u}$ is the desired collection basis functions if and only if $\forall\; x\in\mathbb{R}^{n_x}$ the outputs sequence $\textbf{y}$ driven by $\textbf{u}$, there exist $g\in \mathbb{R}^{n_c}$
\begin{align}\label{eq:condition_lifting}
    \begin{bmatrix}
        Z\\H_L(\textbf{u}_d)\\H_L(\textbf{y}_d)
    \end{bmatrix}g = \begin{bmatrix}
        z\\\textbf{u}\\\textbf{y}
    \end{bmatrix}\;,
\end{align}
where $z:=[\phi_{u,1}(x),\dots,\phi_{u,n_{\phi_u}}(x)]^\top$ and
\begin{align*}
    Z:= \begin{bmatrix}
       \phi_{u,1}(x_0)&\phi_{u,1}(x_1)&\dots&\phi_{u,1}(x_{n_c})\\
        \vdots &\ddots&\ddots&\vdots\\
        \phi_{u,n_{\phi_u}}(x_0)& \phi_{u,n_{\phi_u}}(x_1)&\dots&\phi_{u,n_{\phi_u}}(x_{n_c})\\
    \end{bmatrix}
\end{align*}

\subsection{Leaning the Koopman Basis Functions}\label{sec:learn_koop}
Due to the previous dicussion, the learning of a Koopman operator is converted to learn basis functions that maximizes the satisfaction of the condition~\ref{eq:condition_lifting}. In practice the underlying state for the nonlinear system is not necessarily measured, we therefore make the following assumption
\begin{assumption}\label{ass:meas}
$x_k$ is measurable with respect to the previous $T_{ini}$ step input-output sequence $\{u_i,y_i\}_{i=k-T_{ini}+1}^k$.
\end{assumption}
This assumption implies that $x_k$ can be determined from $\{u_i,y_i\}_{i=k-T_{ini}+1}^k$ and therefore has similar utilization as the matrices $U_p,Y_p$ in problem~\eqref{eq: DeePC} and~\eqref{eq:lin_pred}. Assumed that we have a sequence of input-output data $\textbf{u}_d:=\{u_{d,i}\}_{i=0}^{n_d}$ and $\textbf{y}_d:=\{y_{d,i}\}_{i=0}^{n_d}$ consisting $n_d$ measurements, each of them is partitioned into two subsets, including $\textbf{u}_{d,l}:=\{u_{d,i}\}_{i=0}^{n_{d,t}}$, $\textbf{y}_{d,l}:=\{y_{d,i}\}_{i=0}^{n_{d,t}}$, $\textbf{u}_{d,t}:=\{u_{d,i}\}_{i=n_{d,t}+1}^{n_d}$ and $\textbf{y}_{d,l}:=\{y_{d,i}\}_{i=n_{d,t}+1}^{n_d}$. $n_{d,t}=n_c+T_{ini}+L-1$ is the number of datapoints in the first two sets. The subsets with subscript $_{d,l}$ are used to build the Hankel matrices charactering the Koopman operator while the remaining two subsets are used to learn the basis functions. 

Regarding the Assumption~\ref{ass:meas}, a differentiable learner is used to learn the basis functions, dubbed $\{\phi_{u,\theta}\}_{i=1}^{n_{\phi_u}}$, whose parameters are denoted by $\theta$. Neural networks~\cite{goodfellow2016deep} and Gaussian process~\cite{rasmussen2003gaussian} are recommended learners that have strong representation power. In particular, inducing variables can be considered as trainable parameters for a Gaussian process, please refer to~\cite{titsias2009variational,titsias2010bayesian} for more details. Enforcing the condition~\eqref{eq:condition_lifting} for $L$-step sequences, learning problem is formulated as follows:
\begin{align}\label{eq:learn_koop}
    &\begin{split}
        \min_{\theta} &\sum\limits_{i=n_c}^{n_d} l(\textbf{y}_{d,i}- H_L(\textbf{y}_{d,l})g_i)\\
        \text{s.t.}\;&\; \\
        & g_i = \text{arg}\min_g P(\textbf{u}_{d,i},\textbf{y}_{d,i})\;,
    \end{split}
\end{align}
and 
\begin{align*}
    \begin{split}
    P(\textbf{u}_{d,i},\textbf{y}_{d,i}):& = \lambda_g\norm{g}_2^2 +\lambda_y\norm{Zg-z}_2^2\\
    \text{s.t.}\; &\;  H_L(\textbf{u}_{d,l})g = \textbf{u}_{d,i}\\
    &\; z = \phi_{u,\theta}(\{u_k,y_k\}_{k=i}^{i+T_{ini}-1})
    \end{split}\;.
\end{align*}
In particular, $\textbf{u}_{d,i}:=[u_i,\dots,u_{i+L+T_{ini}-1}]$ and $\textbf{y}_{d,i}:=[y_i,\dots,y_{i+L+T_{ini}-1}], i\geq n_{d,t}+1$ are sequences of inputs and outputs of length $L+T_{ini}$. The matrix $Z$ is the evaluation of the basis functions
\begin{align*}
    Z:= [\phi_{u,\theta}(\{u_i,y_i\}_{i=0}^{T_{ini}}),\dots,\phi_{u,\theta}(\{u_i,y_i\}_{i=n_c-1}^{T_{ini}+n_c-1})]\;.
\end{align*}

The constraint in the learning problem~\eqref{eq:learn_koop} is actually a prediction problem similar to~\eqref{eq:lin_pred} and, therefore, $l(\cdot)$ penalizes the prediction error. As one may notice, there are two relaxations in the learning problem~\eqref{eq:learn_koop}
\begin{enumerate}
    \item To recover an output evaluation, an infinite set of basis functions may be required. This learning problem learns a finite order approximation of these probably infinite set.
    \item The condition~\eqref{eq:condition_lifting} is required to be satisfied for any states, however, the learning problem relax this condition to a set of sampled states. Therefore, the training set $\textbf{u}_{d,t}$ and $\textbf{y}_{d,t}$ should be large enough to represent the condition~\eqref{eq:condition_lifting}.
\end{enumerate}
\begin{figure*}[!htb]
    \centering
    \includegraphics[width = 0.8\textwidth]{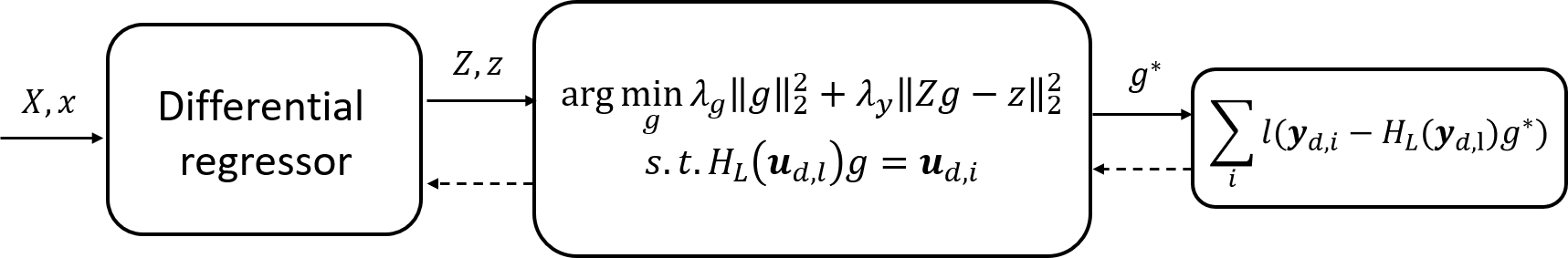}
    \caption{Illustration of learning lifting function framework }
    \label{fig:illustration}
\end{figure*}
Figure~\ref{fig:illustration} shows the flow of the learning problem, where the dashed line indicates the direction of back-propagation. More specifically, the prediction problem is considered as a parametric optimization problem whose differentiation is discussed in Section~\ref{sec:diff_opt}.

Finally, we end this subsection by further listing the benefits of the proposed scheme.
\begin{itemize}
\item Unlike general EDMD methods, which learns the matrices $\mathcal{A,B},C_u$ in~\eqref{eq:lifted_space_state} and~\ref{eq:lifted_sapce_output}. The proposed scheme get rid of the learning of the parameters. Learning of $\mathcal{A,B},C_u$ is ill-conditioned because the solution is not unique.
\item The learning problem optimizes a multi-step forward prediction, meanwhile the utilization of the Willems' fundamental lemma guarantees a good numerical stability in the training scheme, which is a key challenge in training the fully-connected recurrent neural network~\cite{bengio1994learning}. 
\item The proposed scheme is scalable and can be parallelized.
\item Unlike other nonlinear extension of the Willems' fundamental lemma, the proposed scheme does not require nonlinear mapping of future inputs and outputs in the prediction problem.
\end{itemize}

\subsection{Stochastic Prediction}\label{subsec: pred_stoch}
As shown in the Section~\ref{sec:learn_koop}, the prediction problem plays a key role in the proposed scheme. If the chosen learner is deterministic, then the predicted output sequence $\tilde{\textbf{y}}$ driven by inputs $\tilde{\textbf{u}}$ is calculated by
\begin{align}\label{eq:pred_koop}
         \tilde{\textbf{y}} &= H_L(\textbf{y}_{d,l})\tilde{g}&\\
     \tilde{g} &= \text{arg}\min_g&\lambda_g\norm{g}_2^2 +\lambda_y\norm{Zg-z}_2^2\nonumber\\
    &&\text{s.t.}\;\;  H_L(\textbf{u}_{d,l})g = \tilde{\textbf{u}}\nonumber\\
   & &\; z = \phi_{u,\theta}(\tilde{\textbf{u}}_p,\tilde{\textbf{y}_p})\nonumber\;,
\end{align}
which main results in overfitting. Probabilistic learner is one solution to avoid overfitting, such as the aforementioned Gaussian process and the Bayesian neural networks. The output of a probabilistic learner is a distribution but not a deterministic point. In this section, we will show how this distributional outputs from a probabilistic learner can be used for prediction, which is also essential for the training of the Koopman operator. Two methods will be discussed, one is based on Monte-Carlo sampling while another one generates prediction by bounding the Wasserstein distance.

\subsubsection{Monte-Carlo Prediction}
Assuming the distribution of the probabilistic learner is $\mathbb{P}_{\phi_u}$, a Monte-Carlo method is applied to calculate the distribution of the prediction. In particular, the matrix $Z$ in problem~\eqref{eq:pred_koop} is sampled from $\mathbb{P}_{\phi_u}$, which gives a sample from the output distribution. By Monte-Carlo method, the output distribution can be approximated by the sampled outputs.

Meanwhile, the loss function in the learning problem~\eqref{eq:learn_koop} is modified to expected cost. In conclusion, we have the following learning problem
\begin{align*}
    &\begin{split}
        \min_{\theta} &\sum\limits_{i=n_c}^{n_d} \mathbb{E}\;l(\textbf{y}_{d,i}- H_L(\textbf{y}_{d,l})g_i)\\
        \text{s.t.}\;&\; \\
        & g_i \sim \text{arg}\min_g P(\textbf{u}_{d,i},\textbf{y}_{d,i})\;,
    \end{split}
\end{align*}
and 
\begin{align*}
    \begin{split}
    P(\textbf{u}_{d,i},\textbf{y}_{d,i}):& = \lambda_g\norm{g}_2^2 +\lambda_y\norm{Zg-z}_2^2\\
    \text{s.t.}\; &\;  H_L(\textbf{u}_{d,l})g = \textbf{u}_{d,i}\\
    &\; z \sim \mathbb{P}_{\phi_{u,\theta}}(\{u_k,y_k\}_{k=i}^{i+T_{ini}-1})
    \end{split}\;,
\end{align*}
the $k$-th column of the matrix $Z$ follows the distribution $\mathbb{P}_{\phi_{u,\theta}}(\{u_i,y_i\}_{i=k-1}^{T_{ini}+k-2})$. The gradient of this learning problem is also approximated by a Monte-Carlo method.

\subsubsection{Wasserstein Distanced based Prediction}
Intuitively, the regularization term $\norm{Zg-z}_2^2$ in the prediction problem~\eqref{eq:pred_koop} can be considered as the distance between the mean of $Zg$ and $z$. To formulate a more rigor scheme based on a probabilistic learner, we propose to minimize the Wasserstein distance between $Zg$ and $z$. Above all, the entry of  the probabilistic learner output is approximated by a Gausian distribution. Specifically, the $i$-th column of the matrix $Z$ is approximated by $\mathcal{N}(\mu_{i,l},\Sigma_{i,l})$, whose covariance matrix is diagonal
\begin{align*}
    \Sigma_{i,l} = \begin{bmatrix}
        \sigma_{i,1,l}^2&&\\
        &\ddots&\\
        &&\sigma_{i,n_{\phi_u},l}^2
    \end{bmatrix}\;,
\end{align*}
and we denotes the vector composed of the diagonal elements as $\boldsymbol{\sigma}_{i,l}^2$. Accordingly, the $i$-th element of vector $z$ is approximated by $\mathcal{N}(\mu_i,\sigma^2_i)$ and we denote $z\sim\mathcal{N}(\mu_z,\sigma_z)$.

\begin{remark}
\begin{itemize}
    \item It is noteworthy that if an Gaussian process is used to learn the basis function, no approximation is required as the output is already Gaussian.
    \item To better satisfy the condition of a diagonal covariance matrix $\Sigma_{i,l}$, it is recommended to replace the Hankel matrices with Page matrices. In comparison with definition in~\eqref{eq:hankel}, a depth $L$ Page matrix of a sequence $\textbf{w}$ is defined as
    \begin{equation*}
    \mathfrak{P}_L(\textbf{w}) := \begin{bmatrix}
    w_0 & w_L & \dots & w_{(M-1)L}\\
    w_1 & w_{L+1} & \dots & w_{(M-1)L+1}\\
    \vdots & \vdots & \ddots & \vdots \\
    w_{L-1} & w_{2L - 1} & \dots & w_{ML-1}
    \end{bmatrix}\;.
\end{equation*}
\end{itemize}
\end{remark}

Based on the approximation, $Zg$ turns out to be a Gaussian distribution, which is denoted by $Zg\sim \mathcal{N}(\mu_{Zg},\Sigma_{Zg})$ for compactness. To enable a prediction scheme, we conclude the following lemma
\begin{lemma}
The second Wasserstein distance between $Zg$ and $z$ is bounded by 
\begin{align}\label{eq:wass_bound}
    W_2^2(Zg,z)\leq\norm{\mu_{Zg} - \mu_{z}}_2^2 + \norm{\Sigma_{Zg} - \Sigma_{z}}_*
\end{align}
\end{lemma}
\begin{proof}
The second Wasserstein distance between two Gaussian distribution~\cite{villani2008optimal} is
\begin{equation}\label{eq: wasserstein_distance_1}
\begin{aligned}
    &W_2^2(\mathcal{N}(\mu_{Zg}, \Sigma_{Zg}), \mathcal{N}(\mu_z, \Sigma_z)) \\
    = &\norm{\mu_{Zg} - \mu_z}_2^2 + \text{Tr}\left(\Sigma_{Zg} + \Sigma_z\right) \\ - &2\text{Tr}\left(\Sigma_{Zg}^{\frac{1}{2}}\Sigma_z\Sigma_{Zg}^{\frac{1}{2}})\right)
\end{aligned}\;,
\end{equation}
where $\mu_{Zg}= \sum\limits_{i=1}^{n_c}\mu_{i,l}g_i$ with $g_i$ being the $i$-th entry of $g$. The covariance matrix of $Zg$ is calculated as follows
\begin{align*}
    \Sigma_{Zg} = (g^\top\otimes I)\text{Cov}(\text{vec}(Z))(g\otimes I)\;,
\end{align*}
with
\begin{align*}
    \text{Cov}(\text{vec}(Z)) = \begin{bmatrix}
    \Sigma_{1,l} & & \\
     & \ddots & \\
      & & \Sigma_{n_c,l}
    \end{bmatrix}\;,
\end{align*}
therefore, we conclude 
\begin{align}\label{eq: sigma_Zpg}
    \Sigma_{Zg} = \sum\limits_{i=1}^{n_c} g_i^2\Sigma_{i,l}\;.
\end{align}

Since $\Sigma_{z}$ is diagonal, $\Sigma_{Zg}\Sigma_{z} = \Sigma_{z}\Sigma_{Zg}$, \eqref{eq: wasserstein_distance_1} can be reformulated as:
\begin{equation}\label{eq: wasserstein_distance_2}
\begin{aligned}
    &W_2^2(\mathcal{N}(\mu_{Zg}, \Sigma_{Zg}), \mathcal{N}(\mu_{z}, \Sigma_{z})) \\
    = &\norm{\mu_{Zg} - \mu_{z}}_2^2 + \text{Tr}(\Sigma_{Zg} + \Sigma_{z} -2(\Sigma_{Zg}\Sigma_{z})^{1/2})\\
    = &\norm{\mu_{Zg} - \mu_{z}}_2^2 + \text{Tr}((\Sigma_{Zg}^{1/2} - \Sigma_{z}^{1/2})^T(\Sigma_{Zg}^{1/2} - \Sigma_{z}^{1/2}))\\
    = &\norm{\mu_{Zg} - \mu_{z}}_2^2 + \norm{\Sigma_{Zg}^{1/2} - \Sigma_{z}^{1/2}}_F^2
\end{aligned}
\end{equation}
where $\norm{\cdot}_F$ denotes the Frobenius norm. This objective function has a clear interpretation. The first term quantifizes the distance between the mean of these two Gaussian distribution, while the second measures the discrepancy between the covariance matrices. If the derived $\Sigma_{Zg}$ in~\eqref{eq: sigma_Zpg} is substituted in~\eqref{eq: wasserstein_distance_2}, the evaluation of the resulting metric is numerically ill-conditioned.  The Frobenius norm is therefore further relaxed with the Powers-Størmer’s inequality\cite{powers1970free}:
\begin{equation}
    2\text{Tr}(A^\alpha B^{1-\alpha}) \geq \text{Tr}(A + B - |A - B|), 0 \leq \alpha \leq 1
\end{equation}
$A$, $B$ are positive semidefinite and $|A|$ is the positive square root of the matrix $A^*A$, we have:
\begin{equation*}
\begin{aligned}
    &W_2^2(\mathcal{N}(\mu_{Zg}, \Sigma_{Zg}), \mathcal{N}(\mu_{z}, \Sigma_{z})) \\
    \leq &\norm{\mu_{Zg} - \mu_{z}}_2^2 + \text{Tr}(|\Sigma_{Zg} - \Sigma_{z}|)\\
    = &\norm{\mu_{Zg} - \mu_{z}}_2^2 + \norm{\Sigma_{Zg} - \Sigma_{z}}_*
\end{aligned}\;,
\end{equation*}
with $\norm{\cdot}_*$ denoting the nuclear norm.
\end{proof}

Based on this lemma, the prediction problem with a probabilistic learner is reformulated as 
\begin{align}\label{eq:pred_koop_wass}
         \tilde{\textbf{y}} &= H_L(\textbf{y}_{d,l})\tilde{g}&\\
     \tilde{g} &= \text{arg}\min_g&\norm{\mu_{Zg} - \mu_{z}}_2^2 + \norm{\Sigma_{Zg} - \Sigma_{z}}_*\nonumber\\
    &&\text{s.t.}\;\;  H_L(\textbf{u}_{d,l})g = \tilde{\textbf{u}}\nonumber\\
   & &\; z = \phi_{u,\theta}(\tilde{\textbf{u}}_p,\tilde{\textbf{y}}_p)\nonumber\;,
\end{align}

\begin{remark}\label{rmk:huber}
The upper bound~\eqref{eq:wass_bound} is non-smooth, where the absolute value evaluation is ill-condidtioned around $0$~\cite[Chapter 3]{beck2017first}. The absolute value is smoothed by an approach similar to a Huber loss~\cite[Chapter 2]{rockafellar2009variational}, which is defined as:
\begin{equation*}
    L_{\delta}(a) = \begin{cases}
    \frac{1}{2}a^2 & \text{for} |a|<\delta\\
    \delta(|a|-\frac{1}{2}\delta) & \text{otherwise}
    \end{cases}
\end{equation*}
The evaluation of the $i$-th diagonal elements in $|\Sigma_{Zg} - \Sigma_{z}|$ is then reformulated as 
\begin{equation*}
\begin{aligned}
    &(|\Sigma_{Zg} - \Sigma_{z}|)_{ii} \\
    = &\begin{cases}
    \frac{1}{2}(\sigma_{Zg, i}^2 - \sigma_{z, i}^2)^2 & \quad |\sigma_{Zg, i}^2 - \sigma_{z, i}^2|<\delta\\
    \delta(|\sigma_{Zg, i}^2 - \sigma_{z, i}^2|-\frac{1}{2}\delta) & \quad\text{otherwise}
    \end{cases}
\end{aligned}\;.
\end{equation*}
\end{remark}

\section{Koopman-based Data-driven Predictive Control}\label{sec: control}

The DeePC framework (\ref{eq: DeePC}) can be extended into nonlinear systems by integrating the equality constraints with (\ref{eq:condition_lifting}). However, DeePC formulation suffers from the problem that the prediction step interweaves with the control step. In another word, the algorithm may use a non-optimal prediction result for control. When the penalty factor $\lambda_y$ is not sufficient large and the system is initialized away from the reference, the algorithm tends to compensate the difference with a relative large $\sigma_y$, which will result in control failure. To tackle this problem, we propose a bi-level programming formulation\cite{dempe2002foundations}, where the prediction step is independent from control.
\begin{subequations}\label{eq: bilevel formulation}
    \begin{equation}\label{eq: bilevel_control}
    \begin{aligned}
        \min_{\mathbf{u}, \mathbf{y}, g} &(\sum_{k=0}^{N-1}\norm{y_k-r_{t+k}}_Q^2+\norm{u_k}_R^2)\\
        \text{subject to }
        & u_k \in \mathcal{U}, \forall k \in {0, \dotsc, N-1}\\
        & y_k \in \mathcal{Y}, \forall k \in {0, \dotsc, N-1}\\
        &Y_fg = \mathbf{y}\\
        \text{for some } &g \in \Phi(\mathbf{u})\\
    \end{aligned}
    \end{equation}
    \begin{equation}\label{eq: bilevel_prediction}
    \begin{aligned}
        \Phi(\mathbf{u}) = \text{arg}\min_{g}& \lambda_g\norm{g}_2^2+\lambda_y\norm{
        Zg - z}_2^2\\ 
        \text{subject to} & \begin{bmatrix}
        U_p \\ U_f 
        \end{bmatrix} g=
        \begin{bmatrix}
        \mathbf{u}_{ini} \\ \mathbf{u} 
        \end{bmatrix} \\
        \text{with parameter } & z = 
        \phi_{u,\theta}(\textbf{u}_{ini},\textbf{y}_{ini})
    \end{aligned}
    \end{equation}
\end{subequations}
The bi-level problem introduces a hierarchical structure where the upper level problem (\ref{eq: bilevel_control}) indicates the control step and the lower level problem (\ref{eq: bilevel_prediction}) functions as the prediction step. Note that compared to DeePC (\ref{eq: DeePC}), in (\ref{eq: bilevel_prediction}) the squares of the two norm are used so that the objective function of the lower level problem remains smooth. 

A usually used approach to solve a bi-level problem is to transform it into a single level problem. Applying optimality conditions and introducing optimal value function are the two main categories of transformation approaches if the bi-level problem fulfills certain conditions \cite[Chapter 5]{dempe2002foundations}. Here we present the result by replacing the lower level problem (\ref{eq: bilevel_prediction}) with its KKT conditions:
\begin{equation}\label{eq: single level}
\begin{aligned}
    \min_{g, \mathbf{u}, \mathbf{y}, \mu_1, \mu_2} &(\sum_{k=0}^{N-1}\norm{x_k-r_{t+k}}_Q^2+\norm{u_k}_R^2)\\
    \text{subject to} & \begin{bmatrix}
        U_p \\ U_f \\ Y_f
    \end{bmatrix} g=
    \begin{bmatrix}
        \mathbf{u}_{ini}\\ \mathbf{u} \\ \mathbf{y}
    \end{bmatrix} \\
    2g^\top + 2(&Zg - z)^\top Z + \mu_1U_p + \mu_2U_f = 0 \\
    & u_k \in \mathcal{U}, \forall k \in {0, \dotsc, N-1}\\
    & x_k \in \mathcal{X}, \forall k \in {0, \dotsc, N-1}
\end{aligned}
\end{equation}
This equivalent single level problem is solvable by many optimization toolboxes. 

\begin{remark}
Solving a bi-level problem is in general NP-hard. To the best of our knowledge, there is no valid approach to solve a general bi-level optimization problem where the lower level problem is non-convex. It is worth mentioning that the wasserstein distanced based prediction problem presented in Section~\ref{subsec: pred_stoch} is non-convex. We therefore leave the control formulation integrated with wasserstein distanced based prediction as a future work.
\end{remark}

The algorithm is summarized as follows:
\begin{algorithm}[htbp]
\label{alg: Koopman based DeePC}
\SetAlgoLined
    \For{$t=T_{ini},\dotsc$}{
    Set $z = \phi_{u,\theta}(\textbf{u}_{ini},\textbf{y}_{ini})$ \;
    Solve (\ref{eq: single level}) to obtain an optimal input sequence $\boldsymbol{u}^*(0)$ \;
    Set $u_t = \boldsymbol{u}^*(0)$ \;
    Apply $u_t$ to the system and measure $y_t$
    }
    \caption{Koopman based DeePC}
\end{algorithm}

where $\boldsymbol{u}^*(0)$ denotes the first elements of $\boldsymbol{u}^*$.

\section{Simulation results}\label{sec: simulation}
In this section, the prediction results on a Van der Pol oscillator based on Monte-Carlo prediction and wasserstein distanced based prediction are firstly illustrated. Then a numerical experiment of controlling a bilinear motor is presented. We finally demonstrate the potential of our proposed scheme in the large-scale problems with an example of controlling the nonlinear Korteweg-de Vries equation. The source code of the numerical examples can be accessed through \url{https://github.com/RencciW/DataDrivenControlCode}.

\subsection{Stochastic prediction}

We show the results of prediction of trajectories from a Van der Pol oscillator
\begin{equation*}
    \begin{aligned}
        \dot{x} = \begin{bmatrix}
            x_2 \\ \mu(1-x_1^2)x_2-x_1+u
        \end{bmatrix}
    \end{aligned}
\end{equation*}
with $\mu = 1$. we train a 5-layer network (2, 12, 22, 12 and 12) with 1100 data points sampled from 100 random trajectories generated by Van der Pol oscillator. Each layer excluding the input layer is added with a dropout layer. Choose ReLU as activation function for each hidden layer and Adam as optimizer with learning rate $10^{-3}$. Set the dropout rate equals to $0.2$. The code is implemented with PyTorch \cite{paszke2017automatic}. From each of 100 trajectories, we sample 3 trajectory fragments for the construction of Hankel matrix. For a better comparison with following results, in test phase, we choose 3 trajectory fragments from each of 24 trajectories to formulate the Hankel matrix with $T_{ini}=1$ and $N=10$. Test the trained network on 50 data points sampled from trajectories, which are independent from the data used for training and hankel matrix formulation. Forward the same data 120 times to the network and compute the mean value and standard derivation of the prediction. The results are shown below. The light blue color indicates two times the standard derivation.

\begin{figure}[htbp]
    \centering
    \begin{subfigure}[b]{0.45\textwidth}
         \centering
         \includegraphics[width=\textwidth]{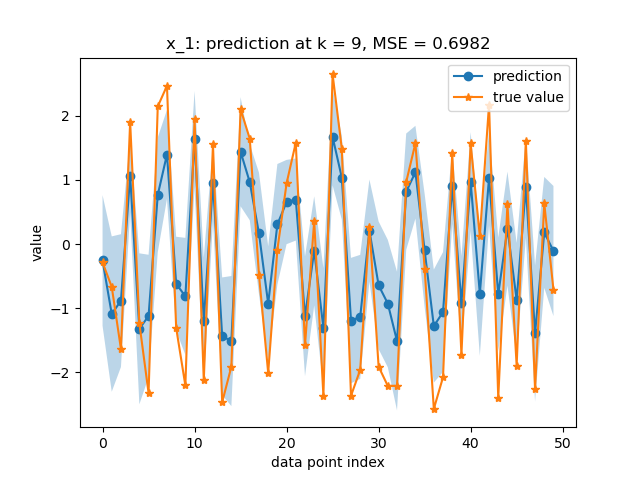}
         \caption{Prediction result ($x_1$) }
     \end{subfigure}
     \vfill
     \begin{subfigure}[b]{0.45\textwidth}
         \centering
         \includegraphics[width=\textwidth]{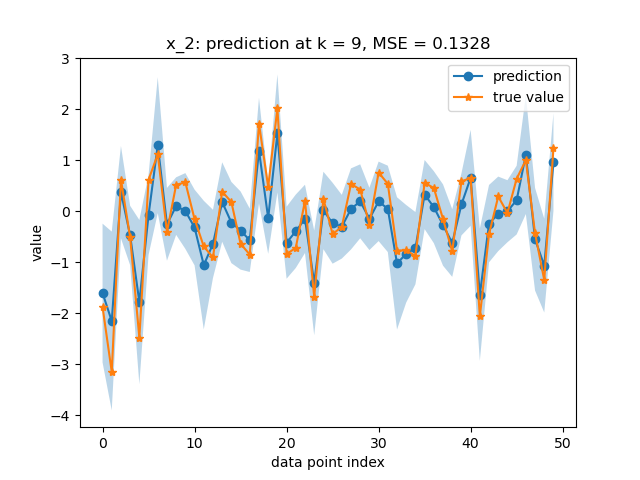}
         \caption{Prediction result ($x_2$)}
     \end{subfigure}
     \caption{Prediction result using Koopman operator learned by MC dropout}
    \label{fig: original_hankel_x}
\end{figure}

\subsection{Prediction based on Wasserstein distance}
We test the new loss function with Van der Pol data. From each of 24 different trajectories, choose one trajectory fragment to formulate the Page matrix. We firstly lift the data for Page matrix construction with the dropout neural network we trained from last subsection. The mean value and the standard derivation of the lifted data are computed for estimation. Then send the data to prediction problem to obtain an optimizer $g^*$. Predict the future trajectory with $x=X_fg^*$, $X_f$ is the Hankel matrix block for prediction.
\begin{figure}[htbp]
    \centering
    \begin{subfigure}[b]{0.45\textwidth}
         \centering
         \includegraphics[width=\textwidth]{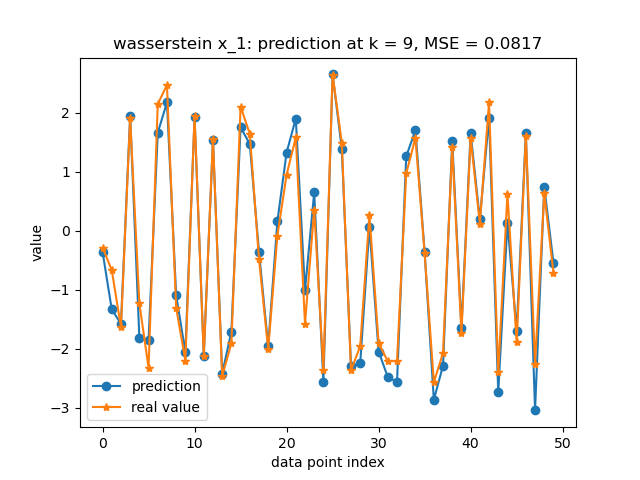}
         \caption{Prediction result $(x_1)$}
     \end{subfigure}
     \vfill
     \begin{subfigure}[b]{0.45\textwidth}
         \centering
         \includegraphics[width=\textwidth]{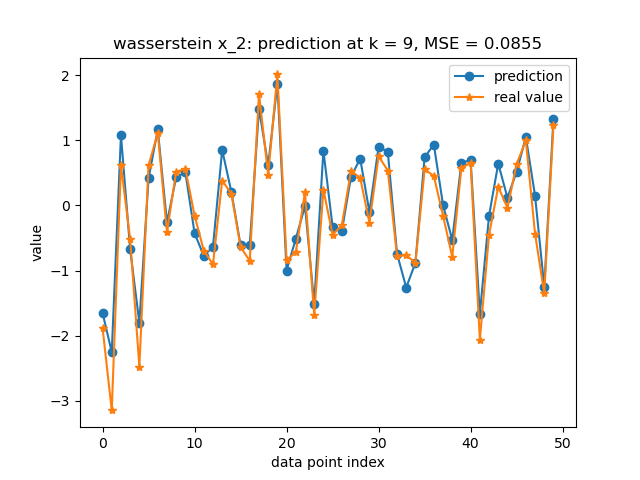}
         \caption{Prediction result $(x_2)$}
     \end{subfigure}
     \caption{Prediction result using loss function derived from Wasserstein distance}
    \label{fig: nuclear_x}
\end{figure}

We compute the mean squared error (MSE) of the prediction based on proposed Wasserstein loss and original quadratic loss at the $9_{th}$ time step. The comparison is summarized in following table:
\begin{table}[!htb]
    \centering
    \begin{tabular}{l|c c}
    \hline
     & $x_1$ & $x_2$\\
    \hline
    Wasserstein loss & 0.0817 & 0.0855\\
    Original loss & 3.8707 & 1.3909\\
    \hline
    \end{tabular}
    \caption{MSE comparison between prediction result computed with different loss functions}
    \label{tab: loss comparison}
\end{table}

It is clear that the Wasserstein loss outperforms the original quadratic loss when uncertainty of lifting function is considered.

\subsection{Control with Koopman-based DeePC}
\subsubsection{Control of a bilinear motor}
We firstly compare the control algorithm  with the algorithm Koopman operator-based MPC controller (K-MPC) proposed in \cite{korda2018linear} by controlling a  bilinear model of a DC motor. \cite{DANIELBERHE1997203}
\begin{equation*}
\begin{aligned}
    \dot{x}_1 &= - (R_a/L_a)x_1 - (k_m/L_a)x_2u + u_a/L_a\\
    \dot{x}_2 &=-(B/J)x_2 + (k_m/J)x_1u-\tau_1/J\\
    y &= x_2
\end{aligned}
\end{equation*}
where $x_1$ is the rotor current, $x_2$ the angular velocity an the control input $u$ is the stator current and the output $y$ is the angular velocity. The parameters are $L_a=0.314, R_a=12.345, k_m=0.253, J = 0.00441, B = 0.00732, \tau_1=1.47, u_a=60$. The physical constraints on the control input are $u\in [-1, 1]$.

We use 40 trajectories with time horizon $0.25s$ to construct a mosaic Hankel matrix. All trajectories are randomly initialized  on the unit box $[-1, 1]^2$. The control input obeys to a uniform distribution over the interval $[-1, 1]$. Choose 40 thin plate spline radial basis function with centers selected randomly with uniform distribution over $[-1, 1]^3$ as lifting functions. Since the system states are not directly measurable, we choose the number of delays $n_d=1$. We define $C =[1, 0, \dotsc, 0]$, $Q= Q_{N_p}=10$, $R=0.01$. The prediction horizon $N = 10$, which implies $0.1s$. Since the system is linear in the lifting space, choose $T_{ini} = 1$. The reference is designed as $r(t) = 0.5cos(2\pi t/3)$. Introduce constraints on output $y \in [-0.4, 0.4]$. 

We simulate for $3s$ and compare the result with a model-based method K-MPC proposed in \cite{korda2018linear}
\begin{figure}[htbp]
\centerline{\includegraphics[width = 0.4\textwidth]{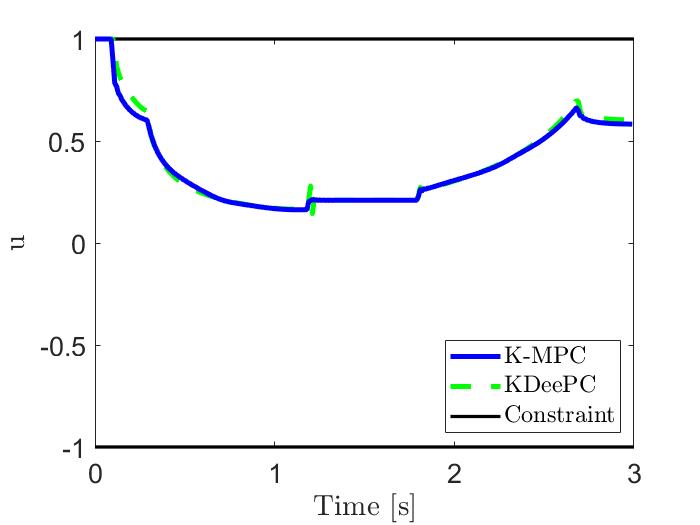}}
\caption{Feedback control input of a bilinear motor}
\label{fig: motor_input}
\end{figure}
\begin{figure}[htbp]
\centerline{\includegraphics[width = 0.4\textwidth]{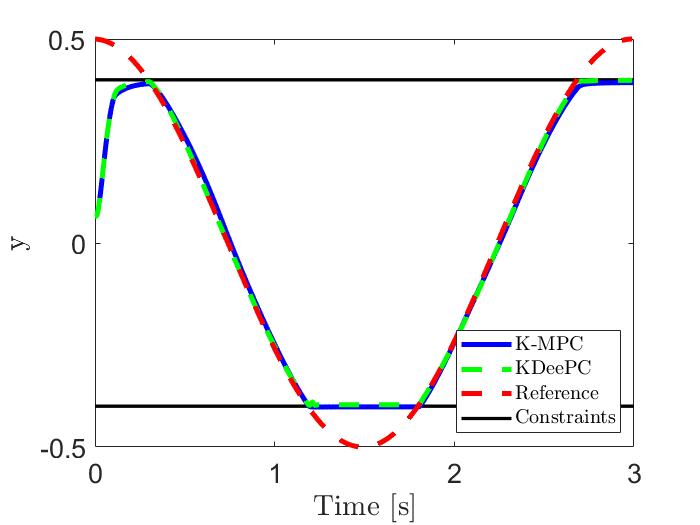}}
\caption{Angular velocity of a bilinear motor}
\label{fig: motor_output}
\end{figure}

As shown in the figure \ref{fig: motor_input} and \ref{fig: motor_output}, the algorithm is capable of following the reference without violating constraints, although compared to K-MPC, the input computed by our method will vibrate gently when the input trajectory is non-smooth.

\subsubsection{Control of nonlinear   Korteweg–de Vries equation}
Our next simulation is to control the nonlinear Korteweg–de Vries(KdV) equation which models the propagation of acoustic waves in aplasma or shallow-water wave \cite{miura1976korteweg}. The equation is given as:
\begin{equation*}
    \frac{\partial y(t, x)}{\partial t} + y(t, x)\frac{\partial y(t, x)}{\partial x} + \frac{\partial ^3y(t, x)}{\partial x^3} = u(t, x)
\end{equation*}
where $y(t, x)$ is the unknown function and $u(t, x)$ the control input. $x\in[-\pi, \pi]$ The space is descretized into 128 points and the time step $\Delta t = 0.02s$. The input is assumed to be of the form $u(t, x)= \sum_{i = 1}^3 u_i(t)v_i(x)$ where $v_i$ consists of 3 spacial basis functions: $v_i(x) = e^{-25(x-\pi/2)^2}$ with $c_1=-\pi/2$, $c_2=0$, $c_3=\pi/2$. The input is constrained to $[-1, 1]$. We initial the system by convexly combining 3 fixed spatial profiles $y_0^1=e^{-(x-\pi/2)^2}$, $y_0^2=-sin^2(x/2)$, $y_0^3=e^{-(x+\pi/2)^2}$. We choose the states itself, the elementwise square of the state, the elementwise product of the states with its periodic shift as the lifting functions. $Q$ is the identity matrix and $R$ is zero matrix. The prediction horizon $N = 5$, which implies $0.1s$. $T_{ini}$ remains equal to $1$. Formulate the Hankel matrix with 63 trajectories, each of which is simulated for $0.5s$.
\begin{figure}[htbp]\label{fig: input_KdV}
\centerline{\includegraphics[width = 0.38\textwidth]{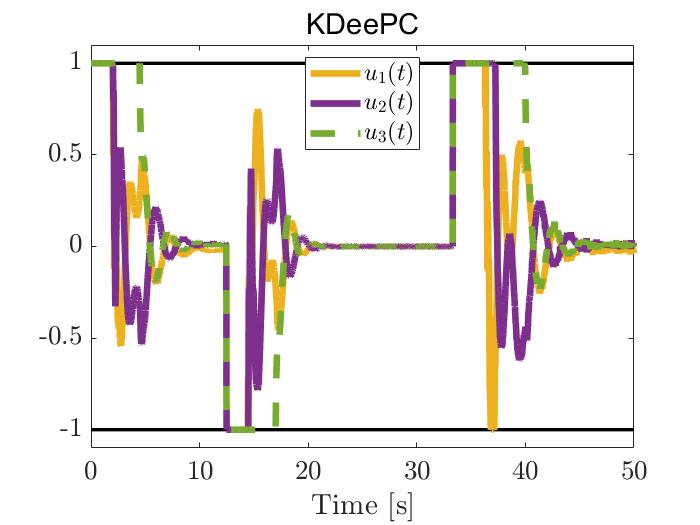}}
\caption{Feedback control input of KdV}
\label{fig: kdV_input}
\end{figure}
\begin{figure}[htbp]\label{fig: KdV_motor}
\centerline{\includegraphics[width = 0.38\textwidth]{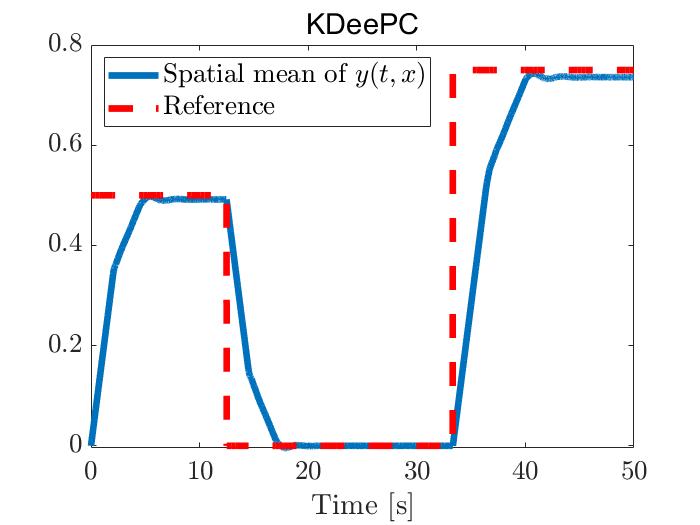}}
\caption{Tracking result}
\label{fig: KdV_output}
\end{figure}

Despite of large dimension, the algorithm is still capable of computing the optimum input in an acceptable time and tracking the reference.

\section{Conclusion}\label{sec: conclusion}

In this work, we extend a data-driven predictive control method into nonlinear systems. The underlying idea is to lift the system with Koopman operator into infinite dimensional space where the system evolves linearly along the nonlinear system trajectories. Approximation of nonlinear lifting functions based on a purely data-driven framework is proposed, along with considerations on the uncertainty of the approximation, which enabling a novel data-driven simulation scheme based on wasserstein distance.


\bibliography{reference}{}
\bibliographystyle{IEEEtran}

\end{document}